\documentclass[10pt,leqno]{amsart}
\usepackage{graphicx}
\usepackage{indentfirst,csquotes}
\usepackage{blindtext}
\usepackage{natbib}

\usepackage{amssymb,amsthm,amsmath}
\usepackage{xcolor,paralist,hyperref,titlesec,fancyhdr,etoolbox}

\titleformat{\section}[display]{\normalfont\huge\bfseries\centering}{\centering\chaptertitlename\thechapter}{10pt}{\Large}
\titlespacing*{\section}{0pt}{0ex}{0ex}

\hypersetup{ colorlinks=true, linkcolor=black, filecolor=black, urlcolor=black }

\usepackage{lipsum}

\begin{document}
\title{Causal Leverage Density: A General Approach to Semantic Information}
\author[Bartlett]{Stuart J Bartlett\\
Division of Geological and Planetary Sciences\\
California Institute of Technology\\
Pasadena, CA 91125\\
United States}
\date{\today}
\address{1200 E California Blvd\\
Pasadena\\
CA 91125}
\email{sjbart@caltech.edu}

\begin{abstract}
I introduce a new approach to semantic information based upon the influence of erasure operations (interventions) upon distributions of a system's future trajectories through its phase space. Semantic (meaningful) information is distinguished from syntactic information by the property of having some intrinsic causal power on the future of a given system. As Shannon famously stated, syntactic information is a simple property of probability distributions (the elementary Shannon expression), or correlations between two subsystems and thus does not tell us anything about the \textit{meaning} of a given message. \citet{kolchinsky2018semantic} introduced a powerful framework for computing semantic information, which employs interventions upon the state of a system (either initial or dynamic) to erase syntactic information that might influence the viability of a subsystem (such as an organism in an environment). In this work I adapt this framework such that rather than using the viability of a subsystem, we simply observe the changes in future trajectories through a system's phase space as a result of informational interventions (erasures or scrambling). This allows for a more general formalisation of semantic information that does not assume a primary role for the viability of a subsystem (to use examples from \citet{kolchinsky2018semantic}, a rock, a hurricane, or a cell). Many systems of interest have a semantic component, such as a neural network, but may not have such an intrinsic connection to viability as living organisms or dissipative structures. Hence this simple approach to semantic information could be applied to any living, non-living or technological system in order to quantify whether a given quantity of syntactic information within it also has semantic or causal power. 
\end{abstract}

\maketitle

Quantifying meaning, or, the meaning of meaning has been as thorny an issue in philosophy as an overgrown rose bush at a balloon party. It's high time this nut was finally cracked. While Shannon and others introduced powerful frameworks for measuring information content, communication channel capacities, mutual information, transfer entropy, etc. \citep{shannon1948mathematical}, it has been notoriously difficult to find precise, mathematical formalisms that can quantify semantic or meaningful information \citep{froese2019problem,kiverstein2022problem,kolchinsky2018semantic,putnam1975meaning}. However, the need for such a formalism is great due to widespread application domains including linguistics, communication, biological semiotics, machine learning, information thermodynamics, control theory, and others. In such disciplines we often need to establish and compute the amount of information in a given system that has causal power over the future of that system. In fact, this is a defining feature of life as compared to non-life: it's incredible power to exert large-scale influence using a relatively small amount of information (e.g., a single strand of DNA) \citep{bartlett2020defining,davies2016hidden,marzen2018optimized,nurse2008life,seoane2018information,zenil2012life}.\par
A primary example of this was the covid-19 pandemic, in which the entire world was plunged into despair as widespread death, suffering and disruption ensued. All of this chaos owed its causal origins to the peculiar informational properties of a handful of molecules that are so small they make a miniscule water droplet seem like an ocean. In this case the virus itself was selected for, or had the emergent `goal' of self-replication and proliferation, which for several years it achieved with ruinous effectiveness. Thus this example is in line with the definition of semantic information introduced in the seminal work of \citet{kolchinsky2018semantic}, whom define it as: ``the syntactic information that a physical system has about its environment which is causally necessary for the system to maintain its own existence''. The information contained in the RNA sequence of the first covid-19 virions to infect humans clearly had significant causal power. On the one hand these molecules contained syntactic information that ensured the viability of many future virus generations. This information also had massive causal influence on the anthroposphere, disrupting the lives of almost every human on earth. In general, it is safe to assume that all similar examples also hold, i.e., the genetic sequence information of biological cells and viruses has significant causal power to support the future viability of the cell, virus or lineage.\par
However, we may want to generalise this idea to situations where the viability of a local subsystem is not the primary variable of interest, where the relevance or quantification of viability is unclear, or situations where we are simply interested in how a quantity of information has causal power over the future unfolding of a system. For example, let us consider the writings and teachings of the Buddha (or the Buddha's subset of the `Dataome' \citep{scharf2021ascent}). This information has had vast causal influence on the world, bringing peace and contentment to the lives of many, perhaps sparing nations from wars, and provoking philosophical debate and inspiration to this day. There is significant `meaning' in this information, but it is not clear whether all of this meaning and causal power is directly linked to the Buddha's viability, or the viability of his lineage (which, according to historical records, probably ceased with his son).\par
We could imagine a hypothetical thought experiment where the Buddha's early inspirations are tragically `scrambled', and he instead grows up as a regular prince instead of the great spiritual master that he in fact was. The world ~2.5 thousand years later would likely look very different, given the global influence of the Buddha's teachings. In the language of statistical physics, the trajectory of the earth system through its accessible phase space would likely have diverged considerably from the trajectory that we are familiar with today. If we assume that in both cases the Buddha's lifespan and the extent of his lineage were the same, then here we have a case of significant meaningful information (the Buddha's teachings), which, when scrambled, has a vast impact on the phase space trajectory of the earth system, while being essentially disconnected from the viability of the subsystem containing that information. A similar logic could apply to any great scholar, scientist, thought leader, etc. They leave a significant legacy of influence upon the world, which in the long term does not have a significant connection to their viability (or in other words, their viability would likely have been similar even in the absence of their influencing the world at large).\par
It seems that even though the viability of the human species may have been enhanced (or at least affected) by such influential figures, the primary variable of interest with respect to semantic information is the future causal influence due to the syntactic information that such people introduced to the world.\par
Another example could be the amino acid sequence of a protein that has been extracted from the last member of a now extinct plant species. Whether this sequence contains meaningful information clearly depends upon whether the functional characteristics of the protein are relevant to the goals and interests of molecular biology and medical engineering. A primary goal of modern medicine is to eliminate cancer cells, for example. How might we objectively quantify the meaningful information of the protein's sequence, while embracing the contextual and contingent features of the esoteric environment that the protein is being studied within (which is of course very different from the environment that it evolved in)? The medical engineers might quantify the efficacy of the protein sequence by the fraction of cancer cells it neutralises. But this is clearly too specific, if we are seeking a definition of semantic information.\par
The unifying feature of this and the previous examples is whether the information at hand (the amino acid sequence), has causal influence on the future unfolding of the wider system the information is embedded within. If the protein turns out to kill all cancer cells that it binds with, then it would revolutionise medicine and the future trajectory of the Anthroposphere, saving millions from death and suffering. Hence the sequence of amino acids in this protein contains a vast quantity of semantic information (or rather it has a high causal leverage density (CLD)), since it would change the course of human evolution, putting us on a very different trajectory to the one where cancer continues killing millions every year. In addition, in physical and informational terms, it is very `small' (a small number of bits and a small physical instantiation).\par
We can even go all the way back to the origins of life on earth. For those who ascribe to an `unlikely event' approach to this process, we can imagine the scenario in Star Trek The Next Generation, where the character Q takes Captain Picard back to the moment when life began on earth. If Q, the prankster that he is, had scrambled the information content of that first protocell or progenote, life would have been literally stopped dead in its tracks, and the earth would have remained sterile for its entire lifespan. Thus in this scenario, unimaginable semantic information (CLD) was present in that progenator pre-lyfeform.\par
Given these examples, and taking significant inspiration from the approach of \citet{kolchinsky2018semantic}, we can formalise the notion of semantic information by comparing phase space trajectories of a given system under interventions that scramble syntactic information. \citet{kolchinsky2018semantic} compare the original system with its intervened counterparts using a metric of viability (whether the subsystem of interest survives or persists). However, in this work I would like to generalise beyond viability by simply comparing the phase space trajectories of the original system with its intervened counterparts. If we assume a known initial state, $R_0$ (point in the system's phase space at time $t=0$), a known complete set of accessible states, and stochastic dynamics (intrinsic uncertainties in the future unfolding of the system's state as described by stochastic differential equations, for example), we can in principle map out a distribution of trajectories that the system could follow as a function of time, weighted by a set of probabilities. We assume that we are interested in some pre-defined time scale $\tau$, for which we allow the system to evolve under its stochastic dynamics.\par
Let us now perform a similar intervention to those suggested in \citet{kolchinsky2018semantic}, i.e., we scramble a portion of the syntactic information in the system. Due to this intervention, the system now has a different initial state, $\hat{R}_0$ (point in the system's accessible phase space). However, the stochastic dynamics remain the same, and we can now allow another set of trajectories to unfold through the system's accessible phase space. It is a set rather than a single trajectory due to the stochastic nature of the dynamics, and after we reach time $t=\tau$, we have a new distribution of final states (points) in phase space, with each point weighted by a probability.\par
At this stage we have two distributions over final states due to the system's stochastic evolution from time $t=0$ to $t=\tau$. The final, un-intervened distribution $p$, will likely be different from its intervened counterpart, $\hat{p}$, but could be very similar or even identical. If $p$ and $\hat{p}$ are similar, that implies the information that was scrambled during the intervention did not have a high semantic content or causal power. However, if the intervened distribution of states is very different from the un-intervened, that implies a large semantic or causal power of the information that was scrambled.\par
In order to quantify this notion, we can use standard tools from information theory to quantify the difference between these two distributions of states. For example, we can employ the symmetric metric for comparing probability distributions, the Jensen-Shannon (J-S) divergence \citep{lin1991divergence}:
\[
D_{JS}(p \| \hat{p}) = \frac{1}{2} D_{KL}(p \| m) + \frac{1}{2} D_{KL}(\hat{p} \| m)
\]

where \( m = \frac{1}{2} (p + \hat{p}) \), and \( D_{KL} \) denotes the Kullback-Leibler divergence, defined as:

\[
D_{KL}(p \| q) = \sum_{x} p(x) \log_2 \left( \frac{p(x)}{q(x)} \right)
\]

Thus, the full expression for the J-S divergence can be written as:

\[
D_{JS}(p \| \hat{p}) = \frac{1}{2} \sum_{x} p(x) \log_2 \left( \frac{p(x)}{m(x)} \right) + \frac{1}{2} \sum_{x} \hat{p}(x) \log_2 \left( \frac{\hat{p}(x)}{m(x)} \right),
\]
where \( m(x) = \frac{1}{2} (p(x) + \hat{p}(x)) \).\par
At this stage we have a measure of the divergence of our un-intervened and intervened systems, capturing the degree to which the information that was scrambled in the intervention has causal power over the evolution of the system. We can now normalise this metric into a dimensionless density, the CLD, by dividing by the number of bits that were erased during the intervention:
\[
\chi_{LD} = \frac{D_{JS}(p \| \hat{p})}{\Omega_{scr}},
\]
where $\Omega_{scr}$ represents the number of bits that were scrambled during the intervention. Hence $\chi_{LD}$ is a dimensionless representation of the causal power that the information erased during the intervention process has on the stochastic evolution of our system over the characteristic time $\tau$. If $\chi_{LD}\approx0$, we can assume that the information erased during the intervention has no causal influence or semantic content on the unfolding of the system's state over the time scale $\tau$. However, large values of $\chi_{LD}$ imply a significant `CLD'. Note that the metric has no upper bound, but is bounded from below at 0 for the case described above.\par
So we now have a general metric for semantic information which encapsulates the idea of causal power on the future evolution of a system due to some quantity of syntactic information within a subsystem. While the metric is dimensionless, being a ratio of two informational quantities with units of bits, it does require the selection of a relevant time scale $\tau$ and the choice of what information-bearing degrees of freedom in the chosen subsystem to randomise. These choices were also addressed by \citet{kolchinsky2018semantic}, and in general it is expected that they will be pre-determined by whomever wishes to compute whether there is semantic information present in a subsystem. For example, biologists would likely target genetic information, and be interested in whether such information has future causal influence on the evolutionary or ecological properties of the broader environment. AI researchers would instead be interested in the weights of their neural networks and the degree to which such information can change the future trajectory of humanity (or perhaps more humble questions).\par
Again, as addressed by \citet{kolchinsky2018semantic}, there may be methods to objectively determine the choice of information to scramble, and characteristic timescale $\tau$, and one could conceive of performing a variational analysis over these quantities in order to \textit{discover} which subsystems contain semantic information. But I leave such questions to future work.\par
In terms of applying this method in the real world, one might suggest that knowing the exact (stochastic) dynamics of most complex systems is difficult at best and impossible at worst. How can we realistically map out the state space of a high-dimensional, non-linear stochastic system, predict ensembles of future phase space trajectories and then analyse the statistics of those distributions of trajectories? Perhaps such a fine-grained level of detail is not strictly necessary to attain useful estimates of CLD. For many systems of interest, we often have robust statistical data (probability distributions) from different instantiations (equivalent to a set of interventions). In this case we can simply apply our expression for $\chi_{LD}$ and still obtain a dimensionless measure of the extent to which our system has diverged due to the erasure of a subset of its information content.\par
Note that in some complex systems, we might expect a compression of the future spread of phase space trajectories due to the destruction of semantic information. For example, let us imagine that an archaeologist unearths an ancient Egyptian text containing a formula for the elixir of life. Clearly this text has significant semantic information, for it could usher in a new path for the Anthroposphere where no human has to die. In fact, depending on stochastic factors, such as whether the text gets inadvertently destroyed by bandits soon after discovery, the future trajectories of humanity now span a spectrum from `business as usual' to `immortal utopia'. If we intervene on the state of the system by destroying the `book of vitality', this broad spread of future trajectories for humanity gets immediately squished into a miniscule  bundle, most of which look fairly mundane compared to the just-erased alternatives. Hence in this scenario, the intervened distribution of trajectories spans a smaller cone than the un-intervened, original distribution. However, this difference will be reflected in the high value of $\chi_{LD}$.\par
The opposite extreme is when the erasure of some semantic information destablises or `blows up' the future set of trajectories in a system's phase space. Such would be the case if some crucial, homeostatic information was destroyed in an intervention. For example, if we were to scramble the genes relevant to cellular gap junctions in a eukaryotic multicellular organism, it's highly likely that cancer would ensue for the unfortunate individual. In this case the intervention has caused a divergence of future trajectories through the system's phase space, from one where the organism maintained cellular homeostasis and integrity (a relatively confined set of future trajectories) to one where a whole range of future paths could now ensue, most of them sadly pathological.\par
There is a conspicuous potential connection between CLD and major transitions in evolution, including the origins of life and the AI revolution. In these transitions, there is a significant jump in the CLD, ushering in a new epoch for life on earth as a result. At the origins of life, molecular information gained causal control over the future infolding of the wider (otherwise abiotic, passive, dead) environment. As we live through the AI revolution, we see the accumulation of semantic information in compressed, neural network representations of human knowledge. Many of these evolutionary transitions involved increases in the viability-based semantic information introduced by \citet{kolchinsky2018semantic} as well as the CLD (often they coincide). However, in many of the major evolutionary transitions, while the viability of the extant life forms increased in the short term, the transitions often ushered in shifts towards novel forms of life in the long term. In the case of AI, it is not clear whether the viability of humans or life as we know it will ultimately improve as a result of the shift, or whether the shift will catalyse the emergence of a new form of lyfe, endowed with the most universal thermodynamic tool of all: CLD.

\vspace{15mm}
\bibliographystyle{plainnat}
\bibliography{main.bib}

\begin{thebibliography}{13}
\providecommand{\natexlab}[1]{#1}
\providecommand{\url}[1]{\texttt{#1}}
\expandafter\ifx\csname urlstyle\endcsname\relax
  \providecommand{\doi}[1]{doi: #1}\else
  \providecommand{\doi}{doi: \begingroup \urlstyle{rm}\Url}\fi

\bibitem[Bartlett and Wong(2020)]{bartlett2020defining}
Stuart Bartlett and Michael~L Wong.
\newblock Defining lyfe in the universe: from three privileged functions to
  four pillars.
\newblock \emph{Life}, 10\penalty0 (4):\penalty0 42, 2020.

\bibitem[Davies and Walker(2016)]{davies2016hidden}
Paul~CW Davies and Sara~Imari Walker.
\newblock The hidden simplicity of biology.
\newblock \emph{Reports on Progress in Physics}, 79\penalty0 (10):\penalty0
  102601, 2016.

\bibitem[Froese and Taguchi(2019)]{froese2019problem}
Tom Froese and Shigeru Taguchi.
\newblock The problem of meaning in ai and robotics: Still with us after all
  these years.
\newblock \emph{Philosophies}, 4\penalty0 (2):\penalty0 14, 2019.

\bibitem[Kiverstein et~al.(2022)Kiverstein, Kirchhoff, and
  Froese]{kiverstein2022problem}
Julian Kiverstein, Michael~D Kirchhoff, and Tom Froese.
\newblock The problem of meaning: The free energy principle and artificial
  agency.
\newblock \emph{Frontiers in Neurorobotics}, 16:\penalty0 844773, 2022.

\bibitem[Kolchinsky and Wolpert(2018)]{kolchinsky2018semantic}
Artemy Kolchinsky and David~H Wolpert.
\newblock Semantic information, autonomous agency and non-equilibrium
  statistical physics.
\newblock \emph{Interface focus}, 8\penalty0 (6):\penalty0 20180041, 2018.

\bibitem[Lin(1991)]{lin1991divergence}
Jianhua Lin.
\newblock Divergence measures based on the shannon entropy.
\newblock \emph{IEEE Transactions on Information theory}, 37\penalty0
  (1):\penalty0 145--151, 1991.

\bibitem[Marzen and Crutchfield(2018)]{marzen2018optimized}
Sarah~E Marzen and James~P Crutchfield.
\newblock Optimized bacteria are environmental prediction engines.
\newblock \emph{Physical Review E}, 98\penalty0 (1):\penalty0 012408, 2018.

\bibitem[Nurse(2008)]{nurse2008life}
Paul Nurse.
\newblock Life, logic and information.
\newblock \emph{Nature}, 454\penalty0 (7203):\penalty0 424--426, 2008.

\bibitem[Putnam(1975)]{putnam1975meaning}
Hilary Putnam.
\newblock \emph{The meaning of" meaning"}.
\newblock University of Minnesota Press, Minneapolis, 1975.

\bibitem[Scharf(2021)]{scharf2021ascent}
Caleb Scharf.
\newblock \emph{The ascent of information: books, bits, genes, machines, and
  life's unending algorithm}.
\newblock Penguin, 2021.

\bibitem[Seoane and Sol{\'e}(2018)]{seoane2018information}
Lu{\'\i}s~F Seoane and Ricard~V Sol{\'e}.
\newblock Information theory, predictability and the emergence of complex life.
\newblock \emph{Royal Society open science}, 5\penalty0 (2):\penalty0 172221,
  2018.

\bibitem[Shannon(1948)]{shannon1948mathematical}
Claude~Elwood Shannon.
\newblock A mathematical theory of communication.
\newblock \emph{The Bell system technical journal}, 27\penalty0 (3):\penalty0
  379--423, 1948.

\bibitem[Zenil et~al.(2012)Zenil, Gershenson, Marshall, and
  Rosenblueth]{zenil2012life}
Hector Zenil, Carlos Gershenson, James~AR Marshall, and David~A Rosenblueth.
\newblock Life as thermodynamic evidence of algorithmic structure in natural
  environments.
\newblock \emph{Entropy}, 14\penalty0 (11):\penalty0 2173--2191, 2012.

\end{thebibliography}

\end{document}